\documentclass[runningheads]{llncs}
\usepackage[T1]{fontenc}
\usepackage{graphicx}

\usepackage{hyperref}
\usepackage{url}
\usepackage{adjustbox}
\usepackage{graphicx}
\usepackage[linesnumbered,ruled,vlined,norelsize]{algorithm2e}
\usepackage{subcaption}
\usepackage{biblatex}
\usepackage{amsmath,amssymb}
\usepackage{booktabs}
\newcommand{\tool}{\textsc{VScan}}
\newcommand{\toolnsat}{\textsc{VScan$_{NSAT}$}}
\newcommand{\toolabc}{\textsc{VScan$_{ABC}$}}
\newcommand{\pgm}{\textsc{Pgm}}
\newcommand{\pgd}{\textsc{Pgd}}
\newcommand{\crown}{\textsc{$\alpha\beta$-Crown}}
\newcommand{\neuralsat}{\textsc{NeuralSAT}}
\newcommand{\marabou}{\textsc{Marabou}}

\newcommand{\pyrat}{\textsc{PyRAT}}
\usepackage{xcolor}
\usepackage{etoolbox}
\usepackage{tcolorbox}
\newtoggle{usecomment}
\settoggle{usecomment}{false}

\makeatletter

\renewcommand{\sectionautorefname}{\S\@gobble}
\renewcommand{\subsectionautorefname}{\S\@gobble}
\renewcommand{\subsubsectionautorefname}{\S\@gobble}
\makeatother

\SetKwInOut{Input}{input}
\SetKwInOut{Output}{output}
\SetKwBlock{Block}{}{}
\DontPrintSemicolon

\begin{document}

\title{Verifying DNN-based Semantic Communication Against Generative Adversarial Noise}

\author{Thanh Le\inst{1} \and Hai Duong\inst{2} \and ThanhVu Nguyen\inst{2} \and Takeshi Matsumura\inst{1}}
\authorrunning{Le et al.}
\institute{National Institute of Information and Communications Technology, Japan \and
George Mason University, USA
% \email{lncs@springer.com}\\
% \url{http://www.springer.com/gp/computer-science/lncs} \and
% ABC Institute, Rupert-Karls-University Heidelberg, Heidelberg, Germany\\
% \email{\{abc,lncs\}@uni-heidelberg.de}
}
\maketitle              % typeset the header of the contribution
\begin{abstract}
    % Aversarial attacks can cause catastrophic failures in safety-critical applications like autonomous vehicles and industrial IoT.\tvn{this sentence has 0 connection to the first. What does safety-critical apps like autonomous vehicles have to do with Semcom?}
    % Semantic communication (SemCom) systems using deep neural networks promise to enhance wireless communication by transmitting only task-relevant semantic features through an autoencoder.
    Safety-critical applications like autonomous vehicles and industrial IoT are adopting semantic communication (SemCom) systems using deep neural networks to reduce bandwidth and increase transmission speed by transmitting only task-relevant semantic features.
    However, adversarial attacks against these DNN-based SemCom systems can cause catastrophic failures by manipulating transmitted semantic features.
    Existing defense mechanisms rely on empirical approaches provide no formal guarantees against the full spectrum of adversarial perturbations.

    We present \tool{}, a neural network verification framework that provides mathematical robustness guarantees by formulating adversarial noise generation as mixed integer programming and verifying end-to-end properties across multiple interconnected networks (encoder, decoder, and task model).
    Our key insight is that realistic adversarial constraints (power limitations and statistical undetectability) can be encoded as logical formulae to enable efficient verification using state-of-the-art DNN verifiers.
    Our evaluation on 600 verification properties characterizing various attacker's capabilities shows \tool{} matches attack methods in finding vulnerabilities while providing formal robustness guarantees for 44\% of properties---a significant achievement given the complexity of multi-network verification.
    Moreover, we reveal a fundamental security-efficiency tradeoff: compact 16-dimensional latent spaces achieve 50\% verified robustness compared to 64-dimensional spaces.
\end{abstract}

\section{Introduction}
\label{sec:intro}

    Next-generation wireless networks are increasingly adopting semantic communication (SemCom) to address the growing demand for intelligent, task-oriented data transmission in applications such as autonomous vehicles, industrial IoT, and augmented reality~\cite{getu2025semantic,yang2023witt,xie2020lite}.
    While traditional communication systems focus on the technical level and treats wireless connectivity as a data pipe without regard for contextual meaning~\cite{yang2022semantic}, SemCom prioritizes understanding the meaning behind transmitted messages and encodes only the necessary information to convey that meaning.
    In recent years, deep learning-based SemCom employs deep neural networks (DNNs) at both transmitter and receiver to transmit task-relevant semantic information, which addresses the evolving requirements of next-generation wireless applications~\cite{sagduyu2024will,getu2025semantic},
    % \tvn{what is it?  why does it need AI?}
    e.g., adaptive compression under varying channel conditions, feature extraction for diverse task types, and real-time semantic understanding.
    Those are capabilities that traditional fixed coding schemes cannot provide but DNNs can learn through end-to-end optimization.
    In particular, DNN-based SemCom jointly trains source/channel coding and modulation alongside a model to perform a specific task. This helps eliminate the need to transmit and reconstruct full message.
    % \tvn{what is semantic communication?}(SemCom)~\cite{yang2022semantic}.

    Goal-oriented SemCom represents a paradigm shift from traditional bit-level transmission to meaning-focused communication, where the system optimizes for task-specific objectives (e.g., ``pedestrian detected 10 meters ahead'') rather than accurately reconstructing the entire transmitted data, achieving up to 90\% bandwidth reduction while preserving safety-critical decision making~\cite{yang2023witt}.
    This approach is particularly crucial for emerging applications, such as autonomous vehicles~\cite{yang2023witt} and industrial IoT~\cite{xie2020lite}, where the ultimate goal is to achieve specific tasks rather than perfect data recovery.
    DNNs excel in this domain~\cite{xie2021deep,peng2025robust,qin2025generative}
    % \tvn{any more recent citations? nothing is going in this topic for the past 4 years?}
    because of their ability to learn complex semantic representations and optimize end-to-end performance across the entire communication pipeline, from source encoding/decoding to channel encoding/decoding and task execution.

    \begin{figure}[t]
        \centering
        \includegraphics[width=0.85\linewidth]{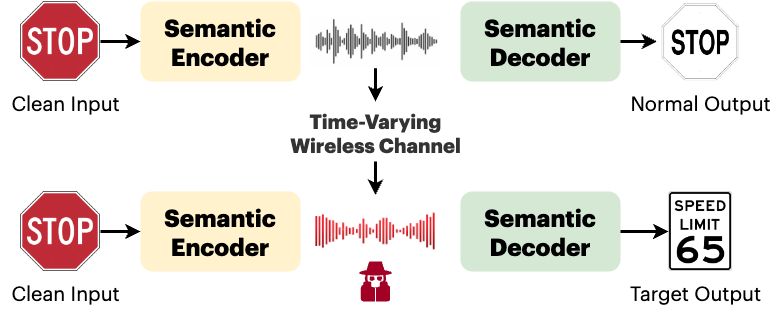}
        \caption{DNN-based SemCom under adversarial noise.}
        \label{fig:abstract}
    \end{figure}

    However, DNN-based SemCom is susceptible to adversarial attacks, in which small perturbations added to the inputs fooling DNNs to misclassify~\cite{bahramali2021robust,li2022sembat,liu2023exploring,wan2025channel,le2025fggm}.
    Such attacks can have severe consequences in SemCom systems, e.g., in autonomous driving scenarios~\cite{yang2023witt}, adversarial perturbations could cause the system to misinterpret traffic signs (\autoref{fig:abstract}) or pedestrians, leading to potentially catastrophic safety failures.
    In industrial IoT applications~\cite{xie2020lite}, these attacks could result in incorrect sensor readings or control decisions, compromising the integrity of critical infrastructure.
    % The semantic nature of these systems makes them particularly vulnerable, as attackers can exploit the learned DNN to craft perturbations that alter the meaning of the encoded feature~\cite{bahramali2021robust}, compromise the task-specific model's outcome, and ultimately facilitate catastrophic consequences.
    The consequences are particularly severe because semantic attacks target the meaning of transmitted data rather than just its reconstruction quality.

    % \tvn{a bit long, shorten the writing}
    %Therefore, it is necessary to develop defense techniques for DNN-based SemCom.
    % \tvn{what does this mean?} before the SemCom or model ensembling~\cite{zhou2024robust}, which further complicates the system\tvn{why?}.

    % \tvn{the limitation of not giving formal guarantees of these 2 classes seems pretty weak -- these techniques might have other benefits (e.g., efficiency) that make them useful and that formal verification cannot provide (i.e., we cannot address).  I would move these into related work and mention that they complement our work rather than being a limitation of prior work.}

    Existing attackers~\cite{madry2017towards,bahramali2021robust,das2021fast} focus on finding specific misclassification examples rather than establishing robustness guarantees across continuous input spaces.
    Recent probabilistic approaches~\cite{feng2025prosac,marzari2024enumerating} provide statistical guarantees for the safety of DNN models through hypothesis testing and safe region enumeration.
    However, these approaches still do not provide formal guarantees for the high-reliability DNN-based SemCom.

    % \tvn{Probably better to just say In this work, we leverage ...}

    In this paper, we present the \emph{Formal \underline{\textbf{V}}erification of \underline{\textbf{S}}emantic \underline{\textbf{C}}ommunication to \underline{\textbf{A}}dversarial \underline{\textbf{N}}oise} (\tool{}) framework, which provides formal guarantees on the robustness of SemCom systems against bounded adversarial perturbations.
    %ereby addressing the fundamental limitation of existing empirical defense methods.
    Our key insight is that realistic adversarial constraints, e.g., power limitations and statistical undetectability requirements, can be encoded as logical formulae to enable efficient formal verification across multiple interconnected networks.
    The core technical challenge addressed by \tool{} is verifying properties across the complete SemCom pipeline from input to final output, e.g.,
    % \tvn{what does this mean?}
    through an encoder, an decoder, and a task-specific model, while handling multiple simultaneous noise sources including adversarial perturbations, input variations, and channel noise.
    % \tvn{too many things going on -- not sure what is the challenging part here?}
    % \tvn{this is challenge because ...  When reading the previous sentence I am not sure which part is hard, e.g., the verification process ?  end-to-end properties, the complete pipeline, the handling of muliple noises ...  . This allows you to connect with the rest}
    This is challenging because traditional DNN verification focuses on single models with single perturbation sources, whereas SemCom requires coordinated analysis across three interconnected networks with multiple interacting noise sources that compound through the pipeline.
    % \tvn{by doing X we were able to leverage}
    By formulating the generative adversarial noise model, e.g., using MIP with sound over-approximated bounds,
    we enable the application of state-of-the-art DNN verification techniques~\cite{wu2024marabou,zhou2024scalable,duong2024harnessing} to verify robustness towards all adversarial noise and input perturbation within a continuous space in a sound and complete manner.
    More importantly, this work introduces SemCom as a new application domain for DNN verification, which is a relatively young field that has primarily focused on traditional adversarial robustness analysis~\cite{brix2024fifth}, and brings mathematical rigor to wireless communication security with implications for safety-critical deployments.

    % \tvn{Given how much space you spent above on intro,  I'd expect a bit more technical details about the approach here. Also, use the name \tool{} more often}
    \tool{} employs a three-phase approach:
    (1) formulating adversarial noise generation as MIP to compute sound overapproximated bounds,
    % \tvn{you already mentioned this above as MIP, so use MIP}
    (2) defining verification properties capturing multiple simultaneous noise sources, and
    (3) leveraging state-of-the-art DNN verifiers to provide formal guarantees across the entire SemCom pipeline from input to classification output,
    % \tvn{again I don't understand what end-to-end formal guarantees means}
    e.g., through all three interconnected networks (encoder, decoder, and pragmatic model).
    By applying \tool{} verification framework specifically to the entire SemCom pipeline, we ensure that the system maintains its intended behavior under adversarial conditions, providing
    % \tvn{this ing connection reads fine}
    necessary assurances for deployment in safety-critical systems.

    % \tvn{I generally don't like using this saying,  "comprehensive" is subjective and invite the reader to really scrutinize the experiments}
    Our evaluation across 600 verification properties and an additional 900 properties in our ablation study demonstrates the effectiveness of \tool{}:
    % \tvn{did you mention this name before?}\hd{only in the abstract}
    (1) \tool{} obtains the same attack capabilities as sophisticated methods like \pgd{} while providing formal
    % \tvn{this is OK}
    robustness guarantees for 263/600 properties that attackers cannot.
    (2) Adversarial noise power constraints
    % \tvn{when I first read this I thought this has something to do with power consumption}
    (limiting the strength of attacks to remain undetectable) significantly impact verification performance, with stricter constraints yielding a higher number of verified properties.
    (3) We identify a fundamental dimensionality trade-off where transmitting lower-dimensional features (16 dimensions) is more secure with 50\% of verified properties.
    In contrast, higher-dimensional spaces (64 dimensions) are more vulnerable, with nearly all of the properties being attacked or undetermined.
    % \tvn{mention why these two things in (3) are important}
    This finding provides concrete design guidance for building secure SemCom systems, which is critical for safety-critical deployments where formal assurances are required.
    % Thus, a power-based adversarial noise detector at the receiver also  in robust SemCom.

    This paper makes the following contributions:
    % \tvn{is this common on Mobicom?}\hd{yes, very common}
    \begin{itemize}
        \item \textbf{Formalized Realistic Threat Model:}
            We formalize realistic adversarial threats against SemCom by capturing practical constraints (input-agnostic perturbations, power limitations, statistical undetectability) in a mathematical framework amenable to DNN verification.
        \item \textbf{Provably Robust SemCom:}
            We compute adversarial noise ranges from \pgm{}~\cite{bahramali2021robust} under power constraints, providing mathematical guarantees for SemCom systems under adversarial conditions.
            \tool{} formally determines when SemCom pipelines are robust against adversarial attacks, providing design principles for secure SemCom systems.
        \item \textbf{End-to-End Verification Framework:}
            \tool{} verifies multiple interconnected networks (encoder, decoder, pragmatic models) with multiple perturbation sources, e.g., adversarial noise, AWGN noise, and input variations.
            This provides end-to-end formal guarantees for complex SemCom systems.
        \item \textbf{Security-Efficiency Tradeoff:}
            \tool{} reveals a fundamental dimensionality principle where compact latent spaces (16) achieve 50\% verified robustness compared to near-zero robustness for high-dimensional spaces (64), providing a concrete design guidance for secure SemCom systems.
        % \item \textbf{Empirical Evaluation:}
            % We conduct experiments on a wide range of factors, including power constraints, latent dimensions, and datasets, demonstrating verification capabilities of \tool{}.
    \end{itemize}

\section{Background}\label{sec:background}
\subsection{DNN-based SemCom}

DNN-based SemCom leverages deep learning's universal function approximation capabilities to enable joint semantic-channel coding~\cite{bourtsoulatze2019deep}.
These systems employ end-to-end architectures with semantic encoders and decoders that extract and reconstruct meaning rather than exact bit sequences~\cite{xie2021deep}, introducing semantic-level metrics that better reflect communication effectiveness.
For example, in autonomous vehicles, instead of transmitting high-resolution images, SemCom transmits concise messages detailing detected elements (e.g., ``pedestrian 10 meters ahead''), significantly reducing bandwidth usage~\cite{yang2023witt}.

\begin{figure}[t]
    \centering
    \includegraphics[width=0.45\linewidth]{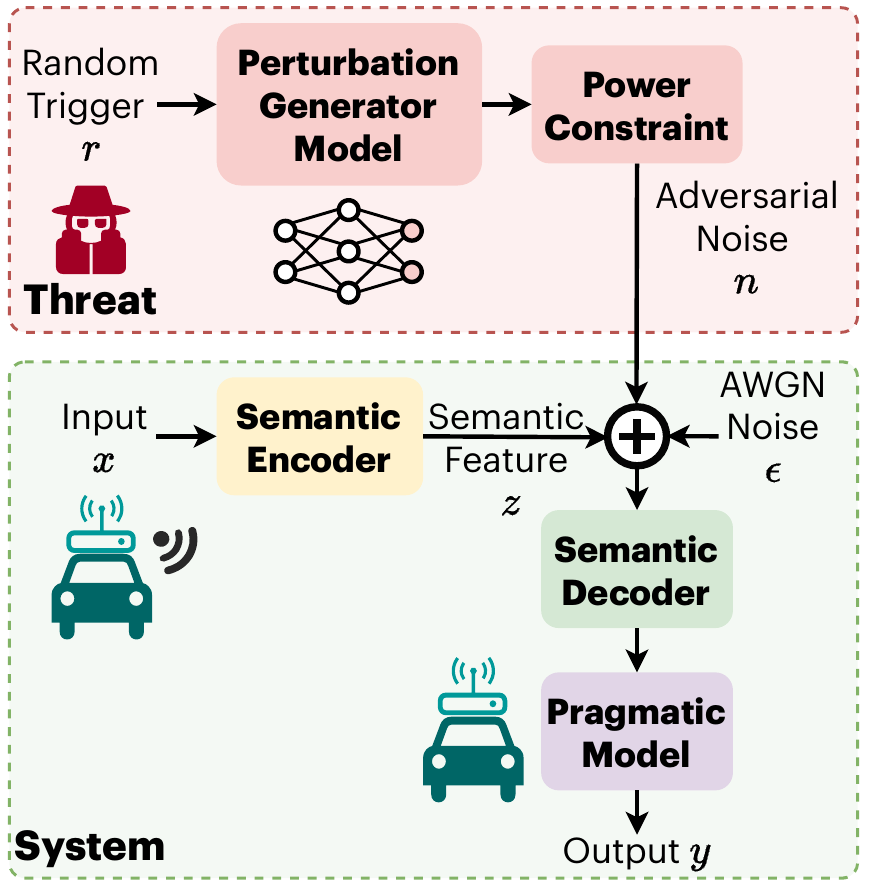}
    \caption{System model and adversarial attacker.}
    \label{fig:threat_model}
\end{figure}

\autoref{fig:threat_model} illustrates a practical DNN-based SemCom consisting of a transmitter (encoder $\mathcal{E}(\cdot)$), a wireless medium, and a receiver (decoder $\mathcal{D}(\cdot)$) \cite{getu2025semantic}.
The transmitter generates a compact semantic feature $z=\mathcal{E}(x)$ from the original data $x$, which is then transmitted over the wireless channel.
Assuming perfect channel estimation, the receiver obtains a noisy signal $z'=z+n$, where $n$ is additive white Gaussian noise (AWGN).
The decoder recovers the message as $x'=\mathcal{D}(z')$, which is fed into a pragmatic model $\mathcal{F}(\cdot)$ to perform downstream tasks, yielding $y=\mathcal{F}(x')$.
The system is then jointly trained: the pragmatic model $\mathcal{F}(\cdot)$ first learns from original data $x$, then the encoder-decoder pair is optimized to extract and reconstruct essential semantic features that maximize $\mathcal{F}(\cdot)$ performance.
This results in semantic features with a smaller footprint compared to traditional communication that aim for perfect reconstruction.

\subsection{Adversarial Attacks}
DNN models are vulnerable to a range of attacks targeting their accuracy \cite{bahramali2021robust,liu2024manipulating}.
% \tvn{citations}\hd{added}.
Adversarial attack techniques aim to find an adversarial input, $\tilde{x}$, such that for a function, $F$, $F(\tilde{x}) \neq F(x)$, where $x$ is the original input.
Several methods exist for generating adversarial samples, e.g.,  Fast Gradient Sign Method (FGSM)~\cite{goodfellow2014explaining} was designed to attack classification models that use Stochastic Gradient Descent~\cite{amari1993backpropagation}.
% FGSM calculates the gradients of the model's loss function with respect to each pixel, then perturbs the input data to maximize the loss function:
% \[\tilde{x} = x + \epsilon \cdot \text{sign}(\nabla_x J(\theta, x, y))\]
Given that FGSM relies on the model's parameters, it is generally considered a white-box attack, indicating the attacker has full access to the model's architecture and parameters.
Note that a white-box attack can be extended to a black-box attack, in which the attacker builds surrogate models by sending queries to obtain input and output pairs from the system~\cite{li2022sembat}.
The Projected Gradient Descent (\pgd{})~\cite{madry2017towards} is a more powerful multi-step variant of FGSM.
\pgd{} operates on a schedule of iterative perturbations where the noise is clipped by a maximum allowed perturbation $\epsilon$ on each iteration.
% \[x(t+1) = \text{clip}_{x, \epsilon} \left( x(t) + \alpha \cdot \text{sign}(\nabla_x J(\theta, x(t), y)) \right)\]
% The \pgd{} process is guided by maximizing the model's loss without accounting for the confidence of the predictions.

Both FGSM and \pgd{} are input-dependent attacks that require knowledge of the specific input being processed to generate targeted adversarial perturbations.
In contrast, the Perturbation Generation Model (\pgm{})~\cite{bahramali2021robust} operates as an input-agnostic attack that generates adversarial noise without knowing the specific semantic features being transmitted.
\pgm{} trains a generative network to produce diverse perturbations that can effectively disrupt SemCom systems across a wide range of inputs.
It is particularly suitable for attacking communication systems where the transmitted content is unknown to the adversary.
% \tvn{what's the purpose of this section?  should add a paragraph at the end mentioning this, e.g., these are good but they lack of formal guarantees which this work addresses}
% \hd{we use some attackers in the evaluation so we want to mention them here. Yes, I will mention about formal guarantees}

\subsection{Formal DNN Verification}
    Given a DNN $N$ and a property $\phi$, the \emph{DNN verification problem} asks if $\phi$ is a valid property of $N$.
    Typically, $\phi$ is a formula of the form $\phi_{in} \implies \phi_{out}$, where $\phi_{in}$ is a property over the inputs of $N$ and $\phi_{out}$ is a property over $N$'s outputs.
    % This form of property has been used to encode safety and security requirements of DNNs, e.g., specifications to avoid collision in unmanned aircraft~\cite{katz2017reluplex}.
    % \tvn{newer citations}

    %  and the counterexample can be used to retrain or debug the DNN~\cite{huang2017safety}.
    % Verification of DNNs using piecewise-linear activation (e.g., ReLU) can be represented as a satisfiability problem~\cite{katz2017reluplex,wu2024marabou,duong2023dpllt,duong2024harnessing}.
    Many DNN verifiers~\cite{katz2017reluplex,wu2024marabou,duong2024harnessing,duong2023dpllt,duong2025neuralsat,duong2025compositional,
    duong2025neuralsat2} treat the DNN verification problem as a satisfiability problem.
    More specifically, given a formula $\alpha$ representing an $L$-layer network $N$ with $N_l$ neurons in layer $l$:
    % \tvn{why ReLU?  can this work support something other than ReLU?}\hd{done}:
    \[
    \alpha \equiv \bigwedge_{\begin{smallmatrix}i \in [1,L]\\j \in [1,N_l]\end{smallmatrix}} v_{i,j} = act \Big( \sum_{k=1}^{N_l} w_{i-1,j,k} \cdot v_{i-1,k} + b_{i,j} \Big)
    \]
    where $act$ is the activation function of the layer (e.g., ReLU, sigmoid, tanh), while $\phi_{in} \implies \phi_{out}$ represents the property to be proved.
    The DNN verification problem then can be reduced to a satisfiability problem~\cite{katz2017reluplex,wu2024marabou,duong2023dpllt,duong2024harnessing,duong2026verifying}:
    \begin{equation}\label{eq:dnn_verification}
        \alpha \land \phi_{in} \land \overline{\phi_{out}}
    \end{equation}
    % A DNN verifier attempts to find a \emph{counterexample} input to $N$ that satisfies $\phi_{in}$ but violates $\phi_{out}$.
    % If no such counterexample exists, $\phi$ is a valid property of $N$; otherwise, $\phi$ is not valid.
    The verifier returns \texttt{unsat} if~\autoref{eq:dnn_verification} is unsatisfiable, indicating that $\phi$ is a valid property of $N$, and \texttt{sat} otherwise, indicating the $\phi$ is not a valid property of $N$~\cite{duong2025generating}.
    % , e.g., a witness that $\phi$ is not valid is a satisfying assignment to the input variables in $\phi_{in}$.

    % \tvn{too short on DNN verification. Talk about some existing work/tools -- what they do, how they work, etc. Here also introduces \neuralsat{}, \crown{} because you'll use them later. Also briefly mention they never considered this semcom setting before.}

    % \tvn{MOVE this pragraph to the end of Sect 2, just above Sect 3 Thread Model}
    State-of-the-art formal DNN verifiers from recent VNN-COMPs~\cite{brix2024fifth,vnncomp2025slides,bak2021second}, such as \crown{}~\cite{zhou2024scalable}, \neuralsat{}~\cite{duong2024harnessing}, \marabou{}~\cite{wu2024marabou}, and \pyrat{}~\cite{lemesle2025verifying},
    % \tvn{not really SOTA anymore, replace with some recent ones like PyRAT}
    employ different forms of branch-and-bound techniques to split verification problems into smaller subproblems and refine bounds through linear relaxations.
    These verifiers leverage CPU/GPU parallelization and advanced optimization techniques to efficiently handle large-scale networks.
    \tool{} leverages \neuralsat{} and \crown{}, the two top tools at VNN-COMP'25~\cite{kaulen20256th} as black-boxes and thus can easily integrate with other DNN verifiers.
    % More broadly, \tool{} provides a new application domain---semantic communication analysis---to DNN verification, a relatively young field that has mostly been focusing on analyzing ML robustness~\cite{brix2024fifth}.\tvn{I like this sentence. Move it somewhere to Intro to indicate novelty of this work is on applicaton.}

    % \tvn{emphasize how this is different than adversarial attacks, otherwise the reader who's not familiar will see they are the same, e.g., both are white box and both aims to find inputs causing violation}\hd{added}.
    % While both formal verification and adversarial attacks may appear similar---both use white-box knowledge and seek inputs that violate model behavior---they serve fundamentally different purposes.
    % Adversarial attacks aim to find \emph{specific} counterexamples that demonstrate vulnerabilities, using gradient-based optimization to identify individual failure cases.
    % In contrast, formal verification provides mathematical guarantees over \emph{continuous input spaces}, either proving that no violating input exists within specified bounds (\texttt{unsat}) or finding a counterexample (\texttt{sat}).
    % Notably, verification offers soundness that when it returns \texttt{unsat}, it formally guarantees that no adversarial attack within the specified constraints can succeed that empirical attack methods fundamentally cannot establish.

    \section{Motivating Example}
    % \tvn{this appears at a strange place. Could you move it before the thread model section? If it requires knowledge of the thread model then make it high level and reference the thread model section. At the end of this section say something: next we talk about formalizing the problem (sect X) and show how to use DNN verification to solve it (sect Y).}
    Consider transmitting an image $x$ (e.g., a red light) through a SemCom system under adversarial conditions, while the pragmatic model is a classifier that classifies the image as red, green, or yellow.
    Suppose that the pramatic model correctly classifies the image as red (e.g., $y_{\text{red}} > y_{\text{green}}$ and $y_{\text{red}} > y_{\text{yellow}}$).
    The system must handle three noise sources simultaneously:
    (1) input perturbations (e.g., blur with strength $s \in [0, 1]$ where $s=0$ is clean and $s=1$ is maximum blur),
    (2) adversarial noise $n$ generated by a trained \pgm{} model (\autoref{sec:threat}) from random trigger $r \in [-1, 1]$, and
    (3) AWGN channel noise $\epsilon \in [-0.01, 0.01]$.

    \tool{} first computes sound over-approximated bounds on adversarial noise by formulating the \pgm{} generator as a mixed integer program (MIP).
    Since the generator involves ReLU activations and quadratic power constraints (\autoref{sec:power_constraint}), \tool{} encodes both the network and the power constraint $\rho$ into the MIP formulation.
    Solving this MIP yields adversarial noise bounds, e.g., $n \in [-0.5, 0.5]$, that over-approximate all possible \pgm{} outputs under the power constraint.

    Next, \tool{} constructs a combined network $N$ representing the complete SemCom pipeline:
    \begin{equation}
        \underbrace{x' = Blur(x, s)}_\text{input perturbation} \land \underbrace{z = \mathcal{E}(x')}_\text{encoding} \land \underbrace{z' = z + n + \epsilon}_\text{channel} \land \underbrace{\hat{x} = \mathcal{D}(z')}_\text{decoding} \land \underbrace{y = \mathcal{F}(\hat{x})}_\text{classification}
    \end{equation}

    The verification property checks whether classification for a red light remains correct under all noise combinations.
    This can be formulated as a satisfiability problem where \tool{} attempts to find a counterexample that satisfies the precondition but violates the property:
    \begin{align*}
        &s \in [0, 1] \land n \in [-0.5, 0.5] \land \epsilon \in [-0.01, 0.01]\\
       \land~ &x' = Blur(x, s) \land z = \mathcal{E}(x') \land z' = z + n + \epsilon \land \hat{x} = \mathcal{D}(z') \land y = \mathcal{F}(\hat{x}) \\
       \land~&(y_{\text{red}} \leq y_{\text{green}} \lor y_{\text{red}} \leq y_{\text{yellow}})
    \end{align*}

    \tool{} invokes state-of-the-art DNN verifiers (e.g., \neuralsat{}) on the combined network and returns \texttt{unsat}.
    This means the system is provably robust: no combination of input perturbations ($s \in [0, 1]$), adversarial noise ($n \in [-0.5, 0.5]$), and channel noise ($\epsilon \in [-0.01, 0.01]$) can cause the pragmatic model to misclassify the transmitted image of a red light into green or yellow.
    % This formal guarantee holds for all possible inputs within the specified bounds, not just the sampled attacks that empirical methods test.

\section{Threat Model}
\label{sec:threat}
% \tvn{Seems more to be background?  I skim through section and don't find it very understandable or important.  So put some bridge building sentences (in plain English:  what it is, why it is important,  at high level stuff ...)  to make each part easier to understand before gettting into the formulae and greek letters.  I would also significantly shorten it as it is too long for background.}
% \hd{design Threat model can be considered as contribution}
% \tvn{I see.  If it's a contribution then clearly say so:  in the intro (e.g., we provide / formalize the model respresenting the threat so that it is ammenable to DNN verification)  and the very first paragraph of that section.   Also make it shorter and less technical (i.e., hide the technical stuff to the appendix or remove them)}

A key contribution of this work is formalizing the adversarial threat against SemCom systems in a way that enables formal verification.
We model realistic adversarial attacks that inject noise into transmitted semantic features to disrupt downstream task performance.
We capture three practical constraints that real attackers face: (1) they cannot predict which specific data will be transmitted, so must generate input-agnostic perturbations; (2) their attack power is limited to avoid detection by simple energy-based monitors; and (3) their noise must be statistically indistinguishable from natural channel noise to evade pattern-based detectors.
By mathematically formalizing these constraints, we transform the threat model into a form amenable to DNN verification techniques, enabling us to provide formal robustness guarantees rather than just empirical defenses.

\subsection{Adversarial Attack on SemCom}

An adversary injects malicious noise into the channel, directly corrupting transmitted semantic features and disrupting the DNN-based SemCom system (\autoref{fig:threat_model}).
We consider a white-box attack where the attacker knows the parameters of both the autoencoder and pragmatic model~\cite{bahramali2021robust},
% \tvn{does this just mean they have the DNN model?}\hd{Yes, attacker knows everything about the models}\hd{Done}
generating adversarial noise to minimize expected system performance across various inputs (\autoref{sec:minimize_expected_performance}).
White-box verification represents the worst-case scenario; if the system withstands this, it is robust against weaker black-box attacks~\cite{li2022sembat}.
We impose practical restrictions (\autoref{sec:power_constraint}, \autoref{sec:statistical_undetectability_constraint}) ensuring the \pgm{} attack remains effective yet undetectable.

\subsection{Minimize Expected Performance}\label{sec:minimize_expected_performance}

Since the attacker cannot predict transmitted data, they must generate input-agnostic perturbations that degrade performance across many inputs.
% \tvn{break into 2 paragraphs}\hd{done}
The adversarial noise generation model $\mathcal{G}_a(\cdot)$ reduces expected performance of decoder $\mathcal{D}(\cdot)$ and pragmatic model $\mathcal{F}(\cdot)$ across a large dataset.
Using \pgm{}~\cite{bahramali2021robust}, the attacker generates perturbation $n = \mathcal{G}_a(r)$ where $r \sim \mathcal{U}(-1, 1)^{|\mathcal{Z}|}$ is a random trigger and $|\mathcal{Z}|$ is the semantic feature dimension.
The training objective maximizes loss on perturbed inputs:
\begin{equation}
    \max_{\mathcal{G}_a} \mathbb{E}_{z, r} \Big[ \ell\Big(\mathcal{F}\big(\mathcal{D}\big(z + \mathcal{G}_a(r)\big)\big), \mathcal{F}\big(\mathcal{D}(z)\big)\Big) \Big],
    \label{eq:pgm_dataset}
\end{equation}
where $\ell$ is the task-specific loss function (e.g., cross-entropy for classification, mean square error for regression).

\subsection{Power Constraint}\label{sec:power_constraint}

Attackers must bound perturbation strength since overly strong noise is easily detected by energy monitors at the receiver.
Unlike prior work using PSNR~\cite{li2022sembat,li2023boosting}, we use peak noise ratio (PNR)
% \tvn{what is it? how is it different than PSNR}
which measures noise-to-signal power ratio in the semantic feature space (where attacks occur) rather than the image space.
We impose the same magnitude limit $\rho$ per dimension during \pgm{} training:
\begin{equation}
    \min_{\mathcal{G}_a} \max\big(\mathcal{G}_a(r)_i^2 - \rho, 0\big) , \quad \forall i
    \label{eq:pgm_power}
\end{equation}
where $\text{PNR (dB)} = 10 * 10 ^{\rho /  \mathbb{E}_{z, i} [|| z_i ||_2^2]}$
% \hd{there is no PNR in the above equation}
and $\mathbb{E}_{z, i} [|| z_i ||_2^2]$ is the expected magnitude of semantic symbols.

\subsection{Statistical Undetectability Constraint}\label{sec:statistical_undetectability_constraint}

Adversarial noise must be statistically indistinguishable from natural AWGN channel noise to evade pattern-based detection.
% \tvn{break into 2 para's}\hd{done}
The \pgm{} model~\cite{bahramali2021robust} uses a GAN architecture~\cite{goodfellow2014generative} where discriminator $\mathcal{D}_{a}$ classifies whether latent space signals are Gaussian distributed.
The generator $\mathcal{G}_a(\cdot)$ is trained to fool the discriminator, matching AWGN noise distribution:
\begin{equation}
    \min_{\mathcal{G}_a} \max_{\mathcal{D}}
        \bigg(\mathbb{E}_{\epsilon}\Big[\log \mathcal{D}_{a}(\epsilon) \Big]
        + \mathbb{E}_{r}\Big[\log \mathcal{D}_{a}\big(\mathcal{G}_a(r)\big) \Big]\bigg),
    \label{eq:pgm_gan}
\end{equation}
where $\epsilon \sim \mathcal{N}^{|\mathcal{Z}|}(0, \sigma_{\text{AWGN}}^2)$ is Gaussian noise with variance $\sigma_{\text{AWGN}}^2$ and $r \sim \mathcal{U}^{|\mathcal{Z}|}(-1, 1)$ is the random trigger.
The complete \pgm{} training loss combines \autoref{eq:pgm_dataset}, \autoref{eq:pgm_power}, and \autoref{eq:pgm_gan}.

\section{The \tool{} Approach}
\label{sec:propose}
    Verifying robustness of DNN-based SemCom systems presents three fundamental challenges that existing empirical approaches cannot address.
    First, SemCom involves multiple interconnected networks (encoder, decoder, pragmatic model) where perturbations propagate through the entire pipeline, requiring coordinated verification across all components rather than isolated analysis.
    Second, realistic operating conditions involve simultaneous noise sources—adversarial perturbations, input variations, and channel noise—that interact in complex ways, demanding verification properties that capture these multi-dimensional uncertainties.
    Third, computing sound bounds on adversarial noise from generative models like \pgm{} requires handling the non-linear, non-convex optimization landscape of neural network generators under practical constraints.

    \tool{} addresses these challenges through a systematic three-phase approach:
    (1) formulating adversarial noise generation as mixed integer programming to compute provably sound over-approximated bounds under power constraints,
    (2) defining comprehensive verification properties that simultaneously handle multiple noise sources across the complete SemCom pipeline, and
    (3) leveraging state-of-the-art DNN verifiers to establish end-to-end formal guarantees that existing methods fundamentally are unable to provide.

    \begin{figure}[t]
        \centering
        \includegraphics[width=\linewidth]{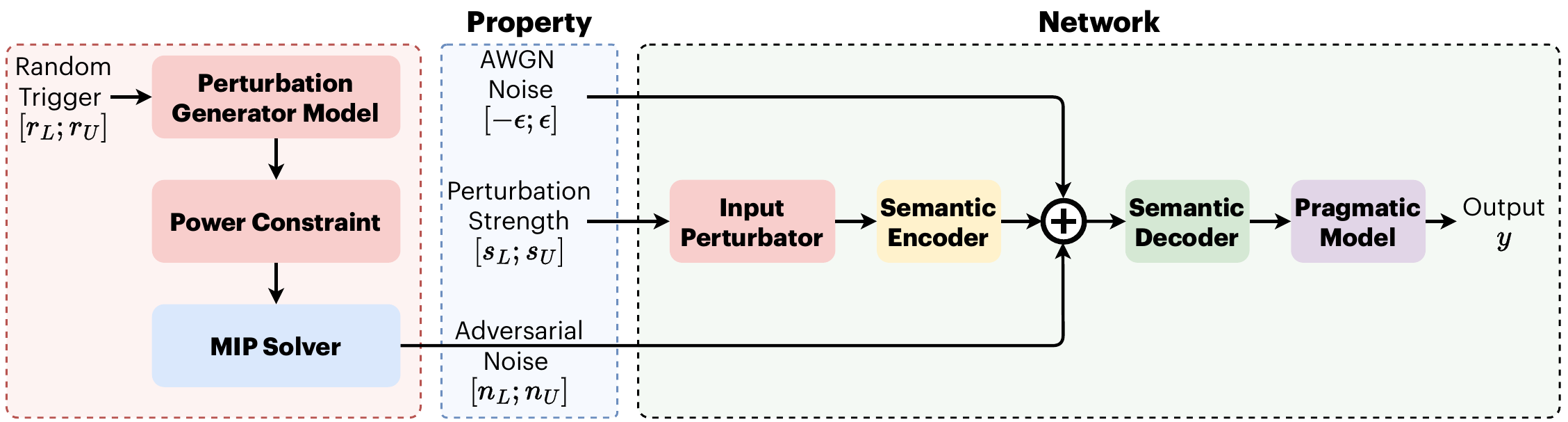}
        \caption{Overview of \tool{}.}
        \label{fig:framework}
    \end{figure}

    \subsection{System Overview}

        % \tvn{There should be some diagram or algorithm showing the flow and interaction between the components. Make it so that the reader can understand the whole process without reading the text.}
        \tool{} provides robustness guarantees for DNN-based SemCom through a three-phase approach (\autoref{fig:framework}).
        First, we formulate the \pgm{} as an MIP to compute over-approximated bounds $[n_L, n_U]$ on adversarial noise $n = \mathcal{G}_a(r)$ under power constraints $\rho$, and
        Next, we define verification properties capturing input variations, adversarial noise, and AWGN channel noise.
        These properties simultaneously handle multiple noise sources and their interactions, which is essential for realistic SemCom operating conditions.
        Finally, we verify that the end-to-end SemCom system (encoder $\mathcal{E}(\cdot)$, decoder $\mathcal{D}(\cdot)$, and pragmatic model $\mathcal{F}(\cdot)$) maintains consistent predictions.
        This phase addresses the unique challenge of verifying multiple interconnected networks where perturbations propagate through the entire SemCom pipeline.
        \tool{} then employs state-of-the-art DNN verifiers to either establish formal guarantees or validate counterexamples through additional MIP solving to distinguish genuine vulnerabilities from spurious results.
        The details of each component are described in the following subsections.

    \subsection{Adversarial Perturbation Space}\label{sec:adversarial_perturbation_space}

        % \tvn{rewrite, break into 2 sentences}\hd{done}
        Computing the adversarial perturbation space generated by the generator model $\mathcal{G}_a(\cdot)$ of \pgm{} involves solving a mixed integer program (MIP) system~\cite{tjeng2019evaluating}.
        This MIP formulation integrates signal power constraints that are specific to our SemCom verification task (\autoref{sec:power_constraint}).

            \begin{equation}
                \begin{aligned}
                    &\mbox{(a)}\quad v^{(i)} = W^{(i)} \hat{v}^{(i-i)} + b^{(i)};~ y = v^{(L)};~  x = \hat{v}^{(0)}; \\
                    &\mbox{(b)}\quad \hat{v}_j^{(i)} \ge {v}_j^{(i)}; \hat{v}_j^{(i)} \ge 0; \\
                    &\mbox{(c)}\quad a_j^{(i)} \in \{ 0, 1\} ;\\
                    &\mbox{(d)}\quad \hat{v}_j^{(i)} \le {a}_j^{(i)} {u}_j^{(i)}; \hat{v}_j^{(i)} \le {v}_j^{(i)} - {l}_j^{(i)}(1 - {a}_j^{(i)}); \\
                    &\mbox{(e)}\quad \big\|y_j\big\|^2_2 \le \rho;
                \end{aligned}
                \label{eq:mip}
            \end{equation}
            where $x$ is input, $y$ is output, and $v^{(i)}$, $\hat{v}^{(i)}$, $W^{(i)}$, and $b^{(i)}$ are the pre-activation, post-activation, weight, and bias vectors for layer $i$.

            These constraints encode DNN semantics, in which
            (a) defines affine transformations and establishes network input/output;
            (b) enforces ReLU activation constraints;% ($\hat{v}_j^{(i)} \ge 0$ and $\hat{v}_j^{(i)} \ge v_j^{(i)}$);
            (c) defines binary activation indicators $a_j^{(i)} \in \{0,1\}$;
            (d) enforces ReLU neuron bounds with upper $u_j^{(i)}$ and lower $l_j^{(i)}$ limits.
            Particularly, deactivating a neuron, $a_j^{(i)} = 0$, simplifies the first of the (d) constraints to $\hat{v}_j^{(i)} \le 0$, and
            activating a neuron simplifies the second to $\hat{v}_j^{(i)} \le v_j^{(i)}$, which is consistent with the
            operations of a ReLU, and
            (e) imposes power limit $\rho$ on \pgm{} outputs.

            Note that the peak power constraint for each output dimension $\|y_j\|_2^2 \leq \rho$ introduces quadratic constraints that standard abstraction-based verifiers cannot directly handle.
            To address this, we effectively employ an MIP solver, e.g., Gurobi~\cite{gurobi}, that supports quadratic programming to calculate these DNN output bounds under these power limitations.
            Note that, the selection of the MIP solver is not a limitation of \tool{}, as any MIP solver that supports quadratic programming can be integrated, e.g., CPLEX~\cite{cplex}, SCIP~\cite{scip}.

    \subsection{Verification Problem}\label{sec:verification_problem}

        Unlike traditional DNN verification, which typically handles single networks with single input specifications, \tool{} requires coordinated verification across the entire pipeline while maintaining formal guarantees.
        \tool{} gathers multiple input properties and connects multiple models in the SemCom pipeline, thereby converting SemCom to a natural setting for DNN verifiers.

        \subsubsection{Verifying Networks:}
        % \hd{perturbation layer}
            \tool{} addresses the unique challenge of verifying an entire SemCom system composed of multiple interconnected DNNs (encoder $\mathcal{E}(\cdot)$, decoder $\mathcal{D}(\cdot)$, and pragmatic model $\mathcal{F}(\cdot)$) under multiple simultaneous input preconditions (adversarial noise $n$, AWGN noise $\epsilon$, and input variations $x$).
            The encoder $\mathcal{E}(\cdot)$ extracts semantic features from the input, and the decoder $\mathcal{D}(\cdot)$ reconstructs semantic features from the received signal. The pragmatic model $\mathcal{F}(\cdot)$ performs the downstream task (i.e., image classification, speech recognition).

        \subsubsection{Verifying Properties:}
            Given a random trigger $r$ and its boundaries $r \in [r_{L}, r_{U}]^{|\mathcal{X}|}$, and power constraint $\rho$, we determine the range of all possibilities of adversarial perturbations $n$ that the \pgm{} can generate by solving the MIP formulation in \autoref{eq:mip}, which yields over-approximated bounds $[n_L, n_U]$ on the adversarial noise.
            Next, for the end-to-end verification, we define the input preconditions for the three components that affect the system.
            % First, to model realistic input perturbations, we employ convolutional perturbations via parameterized kernels~\cite{bruckner2025verification}.
            First, input perturbations $x$ are defined ranging continuously from the lower bound $x_L$ and upper bound $x_U$.
            Second, we determine the adversarial noise bounds $n \in [n_L, n_U]$ from the \pgm{} analysis above.
            Third, we bound the AWGN noise $\epsilon \sim \mathcal{N}(0, \sigma^2_{\text{AWGN}})$ by the 99\% confidence interval, i.e., $\epsilon \in [-3 \times \sigma_{\text{AWGN}}, 3 \times \sigma_{\text{AWGN}}]$.
            % Finally, we define the general input precondition space as:
            % \begin{align}
            %     \mathcal{I} = \{(x, n, \epsilon) :
            %     x \in [x_L, x_U],
            %     n \in [n_L, n_U],
            %     \epsilon \in [-3 \times \sigma_{\text{AWGN}}, 3 \times \sigma_{\text{AWGN}}]\}
            % \end{align}

            However, for high-dimensional inputs such as images, traditional $\ell_p$-norm bounded perturbations assume element-wise independent noise, resulting in computationally intractable verification problems.
            To cope with that, we employ convolutional perturbations using linear parameterized kernels~\cite{bruckner2025verification}, where structured transformations (e.g., blur, sharpen) are parameterized by a single strength variable $s \in [0,1]$, transforming from high into low-dimensional verification problems.
            % We thus refine the input precondition space as:
            Finally, we define the general input precondition space as:
            \begin{align}
                \mathcal{I} = \{(x, s, n, \epsilon) :
                s \in [s_L, s_U],
                n \in [n_L, n_U],
                \epsilon \in [-3 \times \sigma_{\text{AWGN}}, 3 \times \sigma_{\text{AWGN}}]\}
            \end{align}
            where the strength of perturbation $s_L=0$ means there is no perturbation, and $s_U=1$ means the targeted perturbation.
            The verification property (denoted as $\phi$) then asserts that the end-to-end SemCom system's prediction remains robust against both adversarial and channel noise:
            \begin{align}
                &\phi \equiv \forall (x, s, n, \epsilon) \in \mathcal{I} : \mathcal{F}\Big(\mathcal{D}\big(\mathcal{E}(x, s) + n + \epsilon\big)\Big) = \mathcal{F}(x)
            \end{align}
            This property verifies that despite input and latent space variations, the final classification by the complete end-to-end system (encoder, decoder, and pragmatic model) remains consistent with clean input classification.

    \subsection{Verification Framework}
        \autoref{alg:vscan} presents the complete \tool{} verification framework, which systematically establishes formal robustness guarantees for DNN-based SemCom systems through a three-phase approach:
        % \hd{add references to line in the algorithm}
        (1) \emph{Adversarial Noise Computation} constructs an MIP (\autoref{alg:line:mip}) from \pgm{} generator $\mathcal{G}_a$ using \autoref{eq:mip} constraints to compute sound over-approximated bounds $[n_L, n_U]$ on adversarial noise under power constraint $\rho$;
        (2) \emph{Property Construction} (\autoref{alg:line:property}) defines input precondition space $\mathcal{I}$ capturing adversarial noise $n \in [n_L, n_U]$, input variations $s \in [s_L, s_U]$, and AWGN noise $\epsilon$.
        %  then formulates verification property $\phi$ asserting that
        % \[\mathcal{F}(\mathcal{D}(\mathcal{E}(x, s) + n + \epsilon)) = \mathcal{F}(x)  \quad \forall (x, s, n, \epsilon) \in \mathcal{I} \]

        \emph{Adversarial Attack} (~\autoref{alg:line:attack}) attempts to find counterexamples on the composed network $\mathcal{F} \circ \mathcal{D} \circ \mathcal{E}$ and validates them against \pgm{} realizability to eliminate spurious results.
        Since counterexamples are drawn from the over-approximated bounds $[n_L, n_U]$ rather than the exact \pgm{} output space, they can be spurious.
        The validation step performs end-to-end attack from random trigger $r$ through the complete SemCom pipeline to confirm counterexample realizability.
        Finally, \emph{Formal Verification} (\autoref{alg:line:verify}) employs a DNN verifier on the complete system if attack methods fail.
        % , returning \texttt{sat} (vulnerability exists), \texttt{unsat} (formal robustness guarantee), or \texttt{timeout} (computational limit exceeded).
        % \tvn{move this intro of neuralsat and crown to background (see comments there as well as in related work), so that here you just say we use state-of-the-art verifiers such as NeuralSAT and alpha-beta CROWN mentioned in X}
        In particular, we employ two different state-of-the-art complete verifiers,
        % \tvn{why such as?  you did use both of them} \hd{yes}
        \texttt{NeuralSAT}~\cite{duong2024harnessing} and \texttt{$\alpha\beta$-CROWN}~\cite{zhou2024scalable}, to determine the satisfiability of generated verification problems.

        For each instance, the verifier returns one of three possible outcomes:\texttt{sat}, \texttt{unsat}, or \texttt{timeout}.
        %  that have distinct interpretations in the context of SemCom robustness.
        An \texttt{unsat} result verifies that no noise can cause the pragmatic model to produce misclassification, thus certifying the robustness of the SemCom system against the given threat model.
        A \texttt{sat} result indicates that there exists a concrete random trigger $r$ that causes the pragmatic model to produce a different classification.
        A \texttt{timeout} result indicates when the instance exceeds the computational time limit, indicating that the verification problem is too complex for the verifier to solve within the allocated resources.

        \begin{algorithm}[t]
            \caption{\tool{} Verification Framework}
            \label{alg:vscan}
            \Input{
                \pgm{} generator $\mathcal{G}_a$,
                SemCom  $(\mathcal{E}, \mathcal{D}, \mathcal{F})$,
                random trigger $r \in [r_L, r_U]$,
                power constraint $\rho$,
                perturbation $s \in [s_L, s_U]$,
                AWGN $\sigma_{\text{AWGN}}$
            }
            \Output{
                Verification result: \texttt{sat}, \texttt{unsat}, or \texttt{timeout}
            }

            \tcp{Compute Adversarial Noise (\autoref{sec:adversarial_perturbation_space})}
            $m \gets MIP(\mathcal{G}_a, r)$ (\autoref{eq:mip}a-d)\; \label{alg:line:mip}
            \For{$j = 1$ to $|\mathcal{Z}|$}{
                $m.addConstraint(\|n_j\|_2^2 \leq \rho)$ (\autoref{eq:mip}e)\;
            }
            $m.solve()$\;
            $n_{L} \gets m.minimize(n)$\;
            $n_{U} \gets m.maximize(n)$\;
            \tcp{Construct Verification Property (\autoref{sec:verification_problem})}
            $\mathcal{I} \gets \left\{ (x, s, n, \epsilon) \;\middle|\;
                \begin{array}{l}
                    s \in [s_L, s_U], \\
                    n \in [n_L, n_U], \\
                    \epsilon \in [-3 \times \sigma_{\text{AWGN}},\, 3 \times \sigma_{\text{AWGN}}]
                \end{array}
            \right\}$\;

            $\phi \gets \forall (x, s, n, \epsilon) \in \mathcal{I} : \mathcal{F}(\mathcal{D}(\mathcal{E}(x, s) + n + \epsilon))$\; \label{alg:line:property}

            \If(\tcp*[h]{Adversarial Attack}){$attack(\mathcal{F}(\mathcal{D}(\mathcal{E})), \phi)$}{
                \KwRet $validate(\mathcal{F}(\mathcal{D}(\mathcal{E}(\mathcal{G}_a))), r)$\; \label{alg:line:attack}
            }

            \KwRet $verify(\mathcal{F}(\mathcal{D}(\mathcal{E})), \phi)$\label{alg:line:verify} \tcp{Formal Verification}

        \end{algorithm}

\section{Experimental Setups}\label{sec:evaluation}

    \noindent{\textbf{Dataset}.~}
        We use two datasets widely adopted in SemCom literature~\cite{sagduyu2023semantic,nan2023physical,liu2024manipulating}:
        (1) FashionMNIST~\cite{xiao2017fashion} with 28x28 grayscale images across 10 clothing categories; and
        (2) CIFAR10~\cite{krizhevsky2009learning} with 32x32 RGB images across 10 object classes.
        Both datasets compress features to 16-64 dimensions vs. original pixel space while preserving classification accuracy.
        The pragmatic model achieves 88.37\% and 79.85\% accuracy on FashionMNIST and CIFAR10, respectively.

    \vspace{1em}\noindent{\textbf{DNN Hyperparameters}.~}
    % \hd{add more details: layers, learning rate, optimizer}
        The SemCom system comprises an encoder $\mathcal{E}(\cdot)$ (CNN with 8 channels, one FC layer), decoder $\mathcal{D}(\cdot)$ (1-layer FC, transposed CNN), and pragmatic model $\mathcal{F}(\cdot)$ (160k total parameters).
        The encoder compresses inputs into $|\mathcal{Z}|=\{16, 32, 64\}$-dimensional semantic features.
        For pragmatic models, FashionMNIST uses a 2-layer FC network (128 neurons, 10 classes), while CIFAR10 uses ResNet with 3 residual blocks.
        The \pgm{} consists of generator $\mathcal{G}_a(\cdot)$ (3-layer FC, 32 neurons) and discriminator $\mathcal{D}_a(\cdot)$ (2-layer FC, 16 neurons).
        Learning rate is $5 \times 10^{-4}$ for all models.

    % \paragraph{\textbf{Underlying DNN Verifiers}}
    % % \tvn{1 paragraph introducing the two tools, mention crown is ranked 1st and neuralsat is 2nd, etc}\hd{done}
    %     We use state-of-the-art DNN verifiers from VNN-COMP~\cite{brix2024fifth,vnncomp2025slides}: \texttt{$\alpha\beta$-CROWN}~\cite{zhou2024scalable,zhang2022general} and \texttt{NeuralSAT}~\cite{duong2024harnessing,duong2025neuralsat}.
    %     Both leverage GPU-based linear relaxations and Branch-and-Bound (BaB) to efficiently verify DNN properties.

    \vspace{0.5em}\noindent{\textbf{Tools}.~}
        We create two \tool{} variants:
        % such as\tvn{why such as?  you did use both of them} \hd{yes}
        $\toolabc{}$ uses \texttt{$\alpha\beta$-CROWN}~\cite{zhou2024scalable} while
        $\toolnsat{}$ uses {\texttt{NeuralSAT}}~\cite{duong2024harnessing}.
        \tool{} leverages these verifiers as black-box and can work with other DNN verifiers.
        We compare \tool{} against two adversarial attackers:
        % \tvn{could we add (not replace) more recent citations -- saying 2017 or 20121 as state-of-the-art would raise some questions}
        (1) \pgm{}~\cite{bahramali2021robust,liu2024manipulating}, an input-agnostic generative model that produces diverse perturbations without knowing transmitted content; and
        (2) \pgd{}~\cite{madry2017towards,nasr2024projected}, a gradient-based iterative attack using projected gradient descent with complete system knowledge.
        These cover input-agnostic generative attacks to input-specific optimization-based attacks.

    \vspace{1em}\noindent{\textbf{Experimental Platform}.~}
        Our experiments were conducted on a Linux virtual machine in an
        \texttt{a2-highgpu-1g} instance from Google Cloud Platform
        with 12 vCPUs, 85 GB RAM, and a NVIDIA A100 40GB GPU, using Pytorch 2.7.1 and Gurobi 12.0.3.
        % \tvn{would the result be significantly better if you use faster Vcard?}
        % \tl{We will rerun on 4090.}
        % \hd{what's the timeout?}

    \vspace{1em}\noindent{\textbf{Evaluation Benchmarks}.~}
        We use 10 representative images per dataset (one per class).
        % \footnote{\url{https://osf.io/bd56s/?view_only=bf96816e79c54f138ebb3f1ccec7e4c1}}
        For \emph{input properties}, perturbation strength is $s \in [0, 1]$ for FashionMNIST and $s \in [0, 0.5]$ for CIFAR10 using box blur kernels \cite{bruckner2025verification}.
        For \emph{adversarial noise} $[n_L, n_U]$, random trigger $r \in [-1, 1]$ is divided into 10 intervals (e.g., $[-1, -0.8]$ through $[0.8, 1.0]$ for FashionMNIST, $[-0.91, -0.89]$ for CIFAR10).
        \pgm{} models are trained with peak noise ratios $\text{PNR} \in [-10, 10]$ dB, converted to power constraints $\rho$.
        The MIP solver computes adversarial noise intervals $[n_L, n_U]$ from trigger bounds $[r_L, r_U]$ and power constraint $\rho$.

    % \vspace{1em}\noindent{\textbf{Metrics}.~}
    %     We run \pgd{}, \toolnsat{}, and \toolabc{} once per verification instance with 1-minute timeout.
    %     We sample $100$ noise vectors from \pgm{} per instance and report expected \texttt{SAT} instances for stochastic attacks.

    %     Primary metrics include instances returning \texttt{sat} (vulnerability found), \texttt{unsat} (robustness guarantee), or \texttt{timeout} (unknown).
    %     This metric is standard for verification method comparison~\cite{duong2024harnessing,zhou2024scalable,brix2024fifth,bak2021second,vnncomp2025slides}.
    %     We also analyze runtime for computational efficiency across different scenarios.

\section{Results}\label{sec:result}
%     \subsection{Proof-of-concepts software implementation}

%     \tool{} operates as a software-based verification framework that does not require specific hardware implementations or real-time communication infrastructure.
%     The verification process is performed offline on cloud servers, where we check whether the SemCom pipeline satisfies robustness properties under given adversarial conditions before deployment.
%     This design choice enables practical deployment across diverse SemCom hardware platforms without modification, as the formal guarantees established by \tool{} are independent of the underlying transmission medium or physical layer implementation.
%     The verification ensures that messages transmitted through the communication system remain semantically intact and correctly classified by the pragmatic model, even under adversarial perturbations within the verified bounds.

    \subsection{Comparison with Adversarial Attackers}

        \begin{figure}[t]
            \centering
            \begin{subfigure}[t]{0.48\linewidth}
                \centering
                \includegraphics[width=\linewidth]{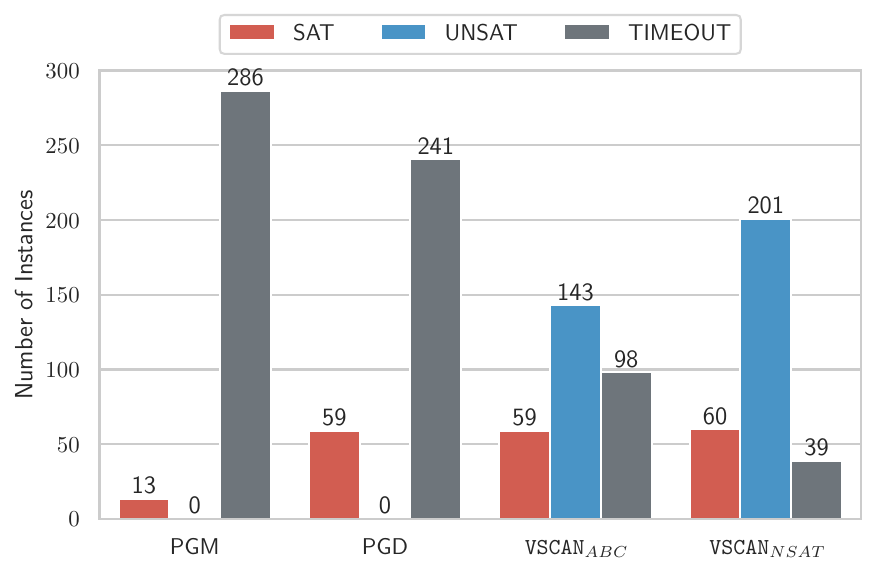}
                \caption{On FashionMNIST benchmark}
            \end{subfigure}
            \begin{subfigure}[t]{0.48\linewidth}
                \centering
                \includegraphics[width=\linewidth]{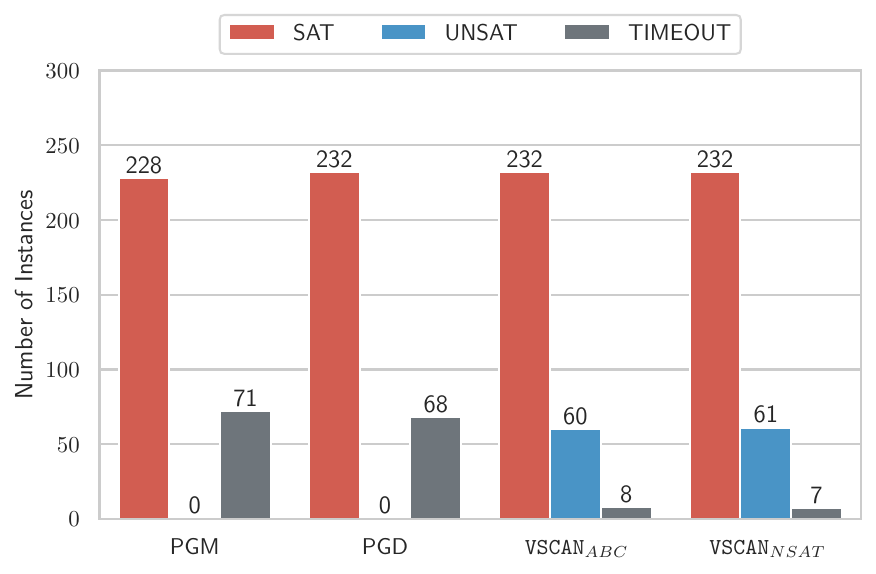}
                \caption{On CIFAR10 benchmark}
            \end{subfigure}
            \caption{Comparison of \tool{} with attackers.}
            \label{fig:bar2}
        \end{figure}

        \autoref{fig:bar2} presents comparisons between \tool{}
        % \tvn{just usetje word \tool{}}
        and existing adversarial attack methods across 600 verification properties
        (300 properties for each dataset, evaluated under PNR values of $-10$, $-7.5$, and $-5$ dB).
        \tool{} demonstrates comparable attacking performance to \pgd{}, e.g.,
        both \toolabc{} and \pgd{} detected 291 vulnerable instances (e.g., 59 from FashionMNIST benchmark and 232 from CIFAR10 benchmark) out of 600 total properties.
        In contrast, \pgm{} demonstrates the lowest attack capability, identifying only 241 vulnerabilities in average (e.g., 13 from FashionMNIST benchmark and 228 from CIFAR10 benchmark), suggesting that \pgm{} may not fully cover the perturbation space.
        % \tvn{NeuralSAT and Crown also use (many) adversarial attacks as the first step,  so you will want to acknowledge and clearly describe the differences here.  Otherwise the reader may think you're hiding something.}
        Note that \pgd{} completely knows the system and its input, while \pgm{} is an input-agnostic attacker, which is more practical.
        \texttt{NeuralSAT} and \texttt{$\alpha\beta$-CROWN} also employ adversarial attacks as their initial step to find counterexamples.

        % While adversarial attack methods can only demonstrate the existence of vulnerabilities through individual adversarial examples, they fundamentally cannot prove their absence.
        When \tool{} fails to find counterexamples, it proceeds to verification using sound mathematical reasoning to formally verify properties (\texttt{unsat}).
        % \tvn{why robustness}\hd{I don't get your question} .
        While attack methods like \pgd{} and \pgm{} can demonstrate the existence of vulnerabilities through individual adversarial examples, they cannot return \texttt{unsat} results, as they fundamentally lack the capability to establish formal robustness guarantees.
        On the other hand, \tool{} provides formal mathematical guarantees
        % \tvn{does it really provide any "proof"?} \hd{no, don't use proof, change to guarantee}
        of robustness for nearly 44\% of the properties, assuring that \emph{no} attack within the specified bounds can compromise the SemCom system, a level of assurance that no empirical attack method can offer.
        This demonstrates that \tool{} provides complete coverage of threat spaces by establishing formal robustness guarantees for properties where attack methods fail.
        % For properties where verification returns \texttt{unsat}, we provide strong mathematical guarantees that no existing adversarial attacks can exploit the system

        \begin{figure}[t]
            \centering
            \begin{subfigure}[b]{0.48\linewidth}
                \centering
                \includegraphics[width=\linewidth]{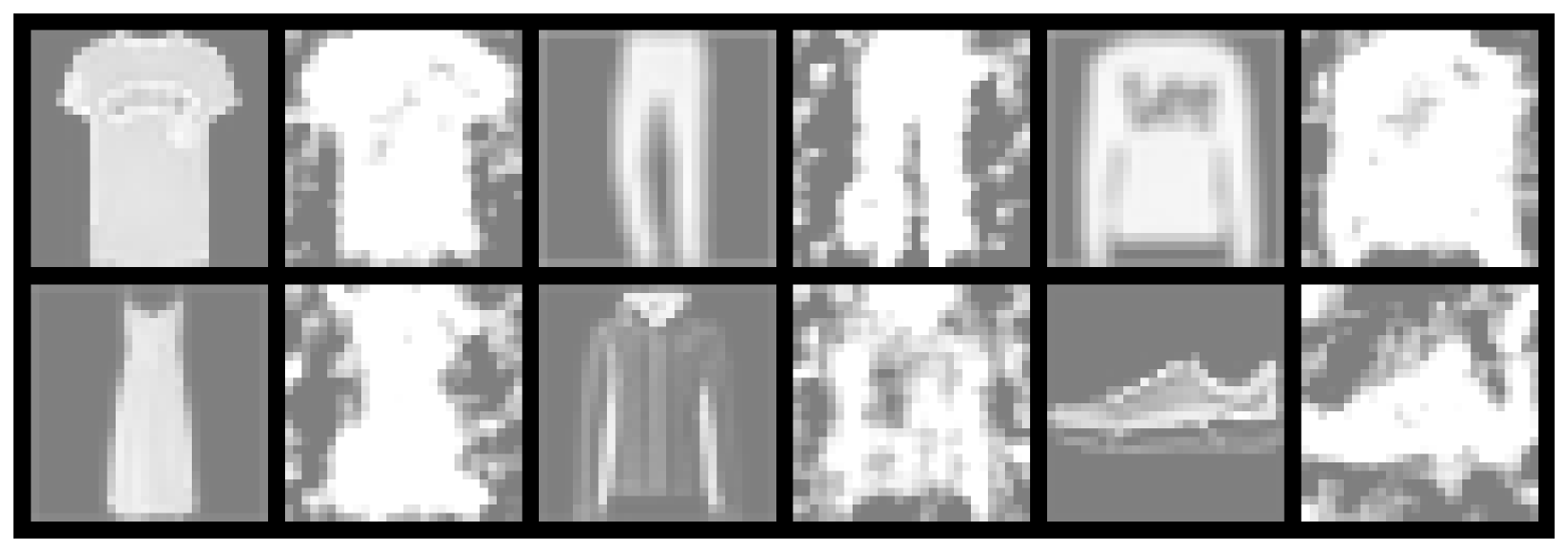}
                \caption{FashionMNIST}
                \label{fig:visualization:fashion-mnist}
            \end{subfigure}
            \begin{subfigure}[b]{0.48\linewidth}
                \centering
                \includegraphics[width=\linewidth]{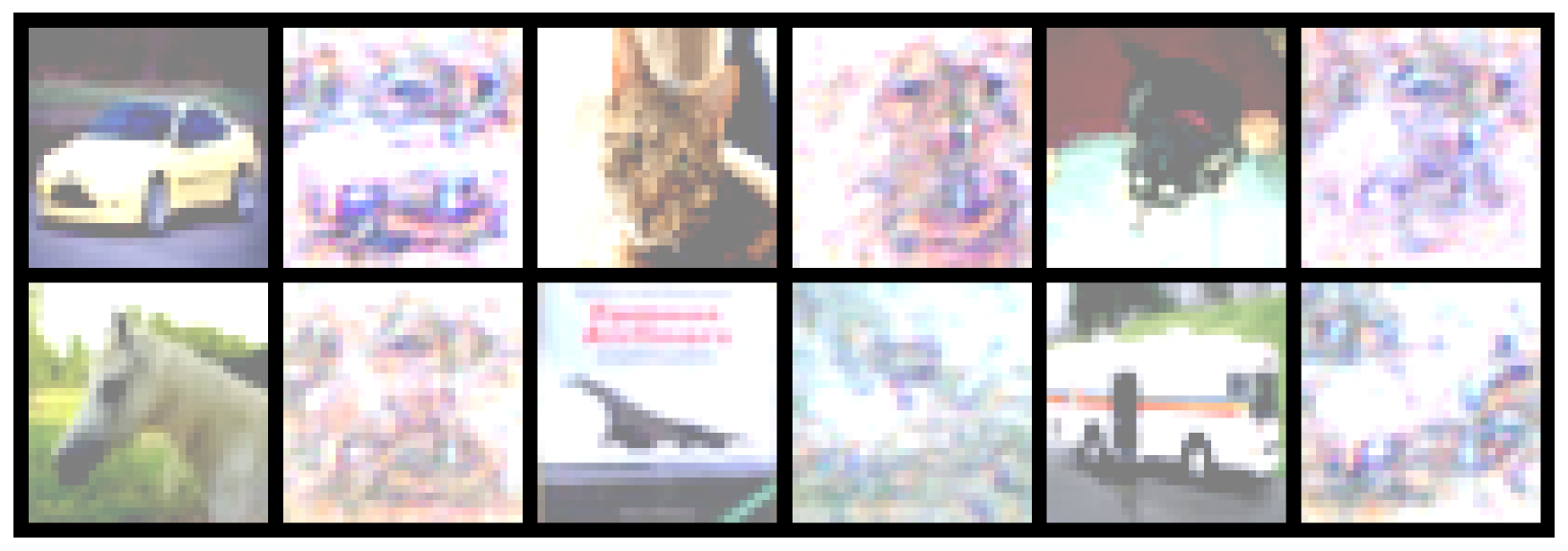}
                \caption{CIFAR10}
                \label{fig:visualization:cifar10}
            \end{subfigure}
           \caption{Examples of clean images and images decoded from perturbed adversarial semantic features, which the pragmatic model correctly classified.}
           \label{fig:visualization}
       \end{figure}

        \autoref{fig:visualization} demonstrates \tool{}'s verification capabilities through concrete examples of images that have been verified as robust (\texttt{unsat}).
        The figure shows pairs of clean images (left) and their corresponding decoded versions (right) after transmission through the SemCom pipeline under adversarial semantic feature perturbations.
        Despite the heavily visible distortions introduced by adversarial noise, the pragmatic model still correctly classifies these decoded images, confirming \tool{}'s formal robustness guarantees.
        These examples illustrate that while the decoded images may appear visually degraded compared to the original clean images, the semantic content essential for the downstream task is preserved within the verified safety bounds established by \tool{}.

    % \begin{table}[t]
    %     \centering
    %     % \setlength{\tabcolsep}{1mm}
    %     \begin{tabular}{cccc}
    %         \toprule
    %         & \multicolumn{3}{c}{\textbf{PNR (dB)}} \\ \cmidrule(l){2-4}
    %         \textbf{Method} & $\mathbf{-5}$ & $\mathbf{0}$    & $\mathbf{5}$  \\ \midrule
    %         \pgm{} (mean)& 26.5/0/273.5& 52.6/0/247.4& 68.9/0/231.1\\
    %         \pgd{}& $65/0/235$ & $104/0/196$ & $123/0/177$ \\
    %         \texttt{\toolabc{}}& $64/64/172$ & $104/28/168$ & $128/20/152$ \\
    %         \texttt{\toolnsat{}}& $66/81/153$ & $103/48/149$ & $125/35/140$ \\
    %         \bottomrule
    %     \end{tabular}
    %     \caption{Solved instances (\texttt{sat}/\texttt{unsat}/\texttt{timeout}).}
    %     \label{tab:result_pnr}
    % \end{table}

    \begin{table}[t]
        \centering
        \setlength{\tabcolsep}{3mm}
        \begin{adjustbox}{width=\linewidth}
        \begin{tabular}{cccc}
            \toprule
            & \multicolumn{3}{c}{\textbf{PNR (dB)}} \\ \cmidrule(l){2-4}
            \textbf{Method} & $\mathbf{-5}$ & $\mathbf{0}$    & $\mathbf{5}$  \\ \midrule
            \pgm{}& 26.5 (23.3) / 0 /273.5 (23.3)& 52.6 (41.3) / 0 /247.4 (41.3)& 68.9 (46.3) / 0 /231.1 (46.3)\\
            \pgd{}& $65/0/235$ & $104/0/196$ & $123/0/177$ \\
            {\toolabc{}}& $64/64/172$ & $104/28/168$ & $128/20/152$ \\
            {\toolnsat{}}& $66/81/153$ & $103/48/149$ & $125/35/140$ \\
            \bottomrule
        \end{tabular}
        \end{adjustbox}
        % \caption{Solved instances (\texttt{sat}/\texttt{unsat}/\texttt{timeout}).}
        \vspace{1em}
        \caption{\tool{} performance on 300 verification problems under different power constraints. Numbers are shown as (\texttt{sat}/\texttt{unsat}/\texttt{timeout}) instances. \pgm{} results show mean (standard deviation) from sampling.}
        \label{tab:result_pnr}
    \end{table}

    \subsection{Performance on Different PNRs}
    \autoref{tab:result_pnr} presents evaluation results across different power constraint levels, measured by PNR in decibels.
    Power constraints significantly impact both attack effectiveness and verification performance across the 300 properties evaluated for each PNR setting.
    The results reveal fundamental tradeoffs between communication power constraints and system security in SemCom systems.

    As PNR increases from -5 to 5 dB, allowing more substantial adversarial perturbations, all methods demonstrate improved attack capabilities.
    At the strictest constraint (PNR = -5 dB), \pgd{} detects 64 vulnerable properties, while both \tool{} verifiers achieve comparable attack detection performance.
    Specifically, \toolabc{} identifies 64 vulnerabilities and \toolnsat{} finds 66, demonstrating that the verification tools' initial attack phases are competitive with specialized adversarial methods.
    At the most relaxed constraint (PNR = 5 dB), sophisticated methods converge to similar performance levels, detecting around 125 vulnerable properties, while \pgm{} reaches only about 69.

    The key distinction lies in \tool{}'s unique capability to provide robustness guarantees that attack methods fundamentally cannot offer.
    \toolnsat{} verified 81 robust properties at the most stringent constraint (PNR = -5 dB), declining to 35 properties as constraints relax (PNR = 5 dB).
    This highlights that tighter power constraints reduces the adversaraial space, thus limit adversarial capabilities and enabling stronger formal guarantees.
    These results reveal a fundamental trade-off where stricter power constraints favor verification tractability by reducing the adversarial search space, while relaxed constraints expand the attack surface and make verification more computationally challenging.
    % \toolabc{} demonstrates a same verification pattern, providing 64 formal guarantees at PNR = -5 dB and decreasing to 20 guarantees at PNR = 5 dB.

    The timeout results also reveal computational complexity patterns across different constraint levels.
    As power constraints relax, both verifiers experience increased timeout rates, with \toolnsat{} showing timeouts decreasing from 153 to 140 properties, while \toolabc{} exhibits a similar trend from 172 to 152 timeouts.
    This suggests that more permissive power constraints create larger and more complex verification search spaces, requiring additional computational resources for complete formal analysis.

    \subsection{Performance on Latent Dimensions}

    % \begin{table}[t]
    %     \centering
    %     % \setlength{\tabcolsep}{1mm}
    %     \begin{tabular}{cccc}
    %         \toprule
    %         & \multicolumn{3}{c}{\textbf{Latent dimension}} \\ \cmidrule(l){2-4}
    %         \textbf{Method} & $\mathbf{16}$    & $\mathbf{32}$   & $\mathbf{64}$  \\ \midrule
    %         \pgm{} (mean)& 75.5/0/224.5& 45.9/0/254.1& 26.5/0/273.5\\
    %         \pgd{}& $106/0/194$ & $95/0/205$ & $91/0/209$ \\
    %         \texttt{\toolabc{}}& $107/98/95$ & $98/13/189$ & $91/1/208$ \\
    %         \texttt{\toolnsat{}}& $105/150/45$ & $95/13/192$ & $94/1/205$ \\
    %         \bottomrule
    %     \end{tabular}
    %     \caption{Solved instances (\texttt{sat}/\texttt{unsat}/\texttt{timeout}).}
    %     % \caption{Tools' performances on 300 verification problems under different latent dimensions. Numbers are shown as (\texttt{sat}/\texttt{unsat}/\texttt{timeout}) instances.}
    %     \label{tab:result_latent}
    % \end{table}

        \begin{table}[t]
            \centering
            \setlength{\tabcolsep}{3mm}
            \begin{adjustbox}{width=\linewidth}
            \begin{tabular}{cccc}
                \toprule
                & \multicolumn{3}{c}{\textbf{Latent dimension}} \\ \cmidrule(l){2-4}
                \textbf{Method} & $\mathbf{16}$    & $\mathbf{32}$   & $\mathbf{64}$  \\ \midrule
                \pgm{}& 75.5 (41.2) / 0 /224.5 (41.2)& 45.9 (37.2) / 0 /254.1 (37.2)& 26.5 (28.9) / 0 /273.5 (28.9)\\
                \pgd{}& $106/0/194$ & $95/0/205$ & $91/0/209$ \\
                {\toolabc{}}& $107/98/95$ & $98/13/189$ & $91/1/208$ \\
                {\toolnsat{}}& $105/150/45$ & $95/13/192$ & $94/1/205$ \\
                \bottomrule
            \end{tabular}
            \end{adjustbox}
            % \caption{Solved instances (\texttt{sat}/\texttt{unsat}/\texttt{timeout}).}
            \vspace{1em}
            \caption{\tool{} performance on 300 verification problems under different latent dimensions. Numbers are shown as (\texttt{sat}/\texttt{unsat}/\texttt{timeout}) instances. \pgm{} results show mean (standard deviation) from sampling.}
            \label{tab:result_latent}
        \end{table}

    \autoref{tab:result_latent} examines how latent space dimensionality affects verification performance of SemCom models. Our results reveal two fundamental patterns.
    First, lower-dimensional latent spaces significantly favor verification procedures.
    At dimension 16, \toolnsat{} establishes robust guarantees for 50\% of the evaluated properties, while \toolabc{} achieves nearly 33\%.
    This verification capability deteriorates drastically as the dimension increases, with \texttt{unsat} results dropping to 13 at dimension 32, and nearly failing at dimension 64 (only a single verified property).
    Second, higher-dimensional latent spaces marginally increase the difficulty in adversarial attack, as
    % favor attack methods by expanding the vulnerable attack surface.
    \pgd{} detects 106 vulnerabilities at dimension 16, 95 at dimension 32, and reaches 91 at dimension 64.

    These findings provide crucial design insights for SemCom systems: \emph{reducing the latent space dimension significantly improves communication robustness by enhancing verification tractability}.
    % \tvn{extend this a bit, make some suggestions to Semcom design or something}
    This dimensionality effect suggests that SemCom should explore compression techniques that preserve semantic fidelity within smaller feature spaces in place of the traditional emphasis on high-dimensional latent representations.
    The results indicate that sacrificing representational capacity for improved verification guarantees may be a worthwhile trade-off for safety-critical applications where formal assurances are essential.

    \subsection{Runtime Analysis}
\begin{figure}[t]
            \centering
            \includegraphics[width=0.45\linewidth]{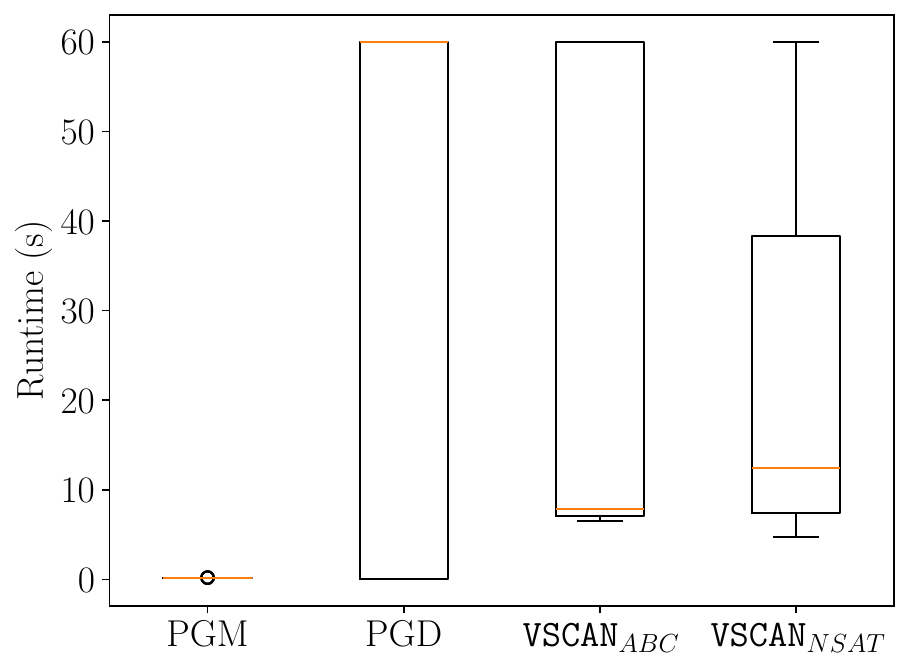}
            \caption{Runtime comparison}
            \label{fig:runtime}
        \end{figure}

        \autoref{fig:runtime} reveals distinct runtime characteristics across different methods that reflect their underlying algorithmic approaches.
        \pgm{} exhibits the fastest performance with a median runtime near 0 seconds, as it involves only single forward passes through the generator network without iterative optimization.
        This efficiency makes \pgm{} highly scalable but comes at the cost of reduced attack effectiveness compared to more input-aware methods.

        In contrast, \pgd{} demonstrates a bimodal runtime distribution with a median around 60 seconds.
        The method achieves fast execution (near 0 seconds) on vulnerable instances where gradient-based optimization quickly finds adversarial examples, but requires the full timeout duration (60 seconds) on robust instances where no successful attack exists.
        This timeout behavior is characteristic of iterative attack methods that lack termination criteria for robust cases.

        The \tool{} variants show intermediate and more predictable runtime patterns.
        \toolabc{} achieves faster performance than \toolnsat{} with median runtimes of approximately 10 seconds and 12 seconds, respectively.
        This difference stems from \toolabc{}'s use of an existing configuration applied for all instances, while \toolnsat{} employs an adaptive configuration selection that automatically chooses the best strategy for each instance.
        Although this adaptive approach increases \toolnsat{}'s runtime overhead, it enables \toolnsat{} to successfully verify more properties.

\section{Discussion}
% \tvn{might not be necessary, but can keep if space allows}

    \tool{} provides formal analysis capabilities that are complementary to, rather than competitive with, existing empirical defense mechanisms.
    Adversarial training techniques~\cite{shi2025secure} and defense methods like denoising autoencoders~\cite{qiu2025plugging} focus on improving system robustness through training-time or inference-time modifications.
    In contrast, \tool{} performs post-hoc analysis of trained models to provide mathematical guarantees about their robustness properties.
    This distinction means that \tool{} can serve as a formal evaluation tool to rigorously assess whether adversarial training or other defense mechanisms have indeed improved model robustness.

    The current \tool{} framework focuses on the semantic encoder-decoder components of SemCom systems, but the underlying approach can be extended to verify additional blocks in the complete communication pipeline.
    For instance, wireless channel encoder-decoder modules that handle modulation and channel coding could be integrated into the verification framework by adding additional DNN components to the end-to-end verification chain.
    This extension would enable formal guarantees across the complete SemCom stack, e.g., from semantic feature extraction to channel coding.
    According to VNN-COMP, DNN verifiers successfully verified networks with size 120M parameters, so it will be aplicable to verify SemCom pipeline with the similar scale.
    However, DNNs used in SemCom are expected to be lightweight for reducing computational delay, models with fewer parameters should be more sufficient for SemCom pipeline.

\section{Related Work}
\label{sec:related}
        % \tvn{You will need to connect these related work with this work. E.g., at the end of the first paragraph on adversaraial attacks, say how this work complements these works by providing a formal verification framework that can prove the robustness of SemCom systems against adversarial attacks.}
        % Recent work has investigated the security of DNN-based SemCom, focusing on adversarial attacks and defenses.
        % Bahramali et al.~\cite{bahramali2021robust} introduced adversarial attacks using generative networks, while Li et al.~\cite{li2022sembat} proposed black-box attacks, and others explored channel-specific vulnerabilities~\cite{wan2025channel,nan2023physical,liu2023exploring}.
        % Defense mechanisms include model ensembling~\cite{zhou2024robust}, adversarial retraining~\cite{shi2025secure}, and denoising autoencoders~\cite{qiu2025plugging}, though these often increase system complexity.

        Recent years have witnessed a surge of interest in the security of deep learning-based semantic communication systems.
        Early work by Bahramali et al.~\cite{bahramali2021robust} introduced robust adversarial attacks against DNN-based wireless communication, highlighting the vulnerability of such systems to perturbations generated by a well-trained generative network.
        Li et al.~\cite{li2022sembat} extended this line of research by proposing black-box physical layer attacks, leveraging surrogate models and OFDM processing to enhance attack effectiveness.
        Other works have investigated the impact of channel statistics~\cite{wan2025channel} and physical-layer characteristics~\cite{nan2023physical} on both the efficacy of attacks and defenses.
        Hardware implementation on universal software radio peripheral and practical vulnerabilities have also been demonstrated~\cite{liu2023exploring}, underscoring the real-world relevance of these security concerns.

        To defend against adversaraial attacks, DNN-based SemCom often incorporates denoising techniques, e.g., autoencoders~\cite{qiu2025plugging}, or model ensembling~\cite{zhou2024robust}.
        %  which further complicates the system\tvn{why?}.
        However, these techniques significantly increase system complexity, e.g., running denoising steps to remove adversaraial noise before the SemCom.
        More importantly, they do not provide formal guarantees against the full spectrum of adversarial perturbations.
        Another popular mechanism is adversarial (re)training~\cite{shi2025secure}, which enhances the robustness of the target system. Yet, there is no formal guarantee that such defenses can prevent unseen attacks.
        More general, such defense mechanisms are developed based on a limited dataset of adversarial samples and leave systems vulnerable to perturbations outside the training set.
        Our work complements these approaches by providing a formal verification framework that mathematically proves SemCom's robustness against adversarial attacks.

        % \tvn{be careful here, this seems to be a common and acceptable limitation. No solution is expected to provide comprehensive threat coverage, and mentioning this will make the reader expect your approach to be able to provide a comprehensive treatment. Soften the language}.

\section{Conclusion}
\label{sec:conclusion}
% \tvn{A bit too long and also reads like a summary. The difference between summary and conclusion is that summary is for people who have not read the paper, while conclusion is for people who have read the paper. So conclusion should just restate the key contributions and future direction.}

    % \tvn{the first 2 sentences are summary}
    % DNN-based SemCom systems promise to revolutionize wireless networks through intelligent, task-oriented transmission, but their susceptibility to adversarial attacks poses critical threats to safety-critical applications.
    % Existing empirical defense mechanisms provide no formal guarantees against adversarial perturbations.
    % This fundamental limitation necessitates formal verification approaches that can provide mathematical assurances of system robustness.
    \tool{} establishes a formal verification framework for DNN-based SemCom, moving beyond empirical evaluation to provide mathematical robustness guarantees.
    The key contribution lies in demonstrating that formal reasoning can provide complete coverage of threat spaces while matching empirical attack methods in finding vulnerabilities.
    % \tvn{try to rewrite to remove new line with single word}

    % \tvn{if these 3 fundable insights are NOT already in the paper, e.g., in the intro or emphasized in evaluation, then you should put them there.  And if they are already there, then you should not mention them here. Conclusion is not a place to introduce new insights.}
    Our evaluation demonstrates three key results:
    (1) \tool{} matches attack methods in finding vulnerabilities while providing formal robustness guarantees for 44\% of properties where attackers fail,
    (2) power constraints create a fundamental trade-off where stricter constraints (PNR = -5 dB) enable 81 verified properties versus 35 under relaxed constraints (PNR = 5 dB), and
    (3) latent dimensionality critically impacts robustness with 16-dimensional spaces achieving 50\% verified properties compared to near-zero for 64-dimensional spaces.

    \clearpage

    \printbibliography

\end{document}